\documentclass[a4paper]{jpconf}
\usepackage{graphicx}
\usepackage{amsmath}
\usepackage{lineno}
\newcommand{\fsm}{\ensuremath{f^{\textrm{SM}}}}
\newcommand{\dphi}{\ensuremath{\Delta \phi}}
\newcommand{\dphidQ}{\ensuremath{\Delta \phi(l, d)}}
\newcommand{\dphibQ}{\ensuremath{\Delta \phi(l, b)}}
\newcommand{\ttbar}{\ensuremath{t\bar{t}}}
\def\com{center-of-mass}
\def\TeV{\ifmmode {\mathrm{\ Te\kern -0.1em V}}\else
                   \textrm{Te\kern -0.1em V}\fi}%
\def\GeV{\ifmmode {\mathrm{\ Ge\kern -0.1em V}}\else
                   \textrm{Ge\kern -0.1em V}\fi}%
\def\MeV{\ifmmode {\mathrm{\ Me\kern -0.1em V}}\else
                   \textrm{Me\kern -0.1em V}\fi}%
\def\keV{\ifmmode {\mathrm{\ ke\kern -0.1em V}}\else
                   \textrm{ke\kern -0.1em V}\fi}%
\def\eV{\ifmmode  {\mathrm{\ e\kern -0.1em V}}\else
                   \textrm{e\kern -0.1em V}\fi}%

\newcommand{\intlumi}{$4.6$~fb$^{-1}$}

\newcommand{\ejets}{\ensuremath{e+\textrm{jets}}}
\newcommand{\mujets}{\ensuremath{\mu+\textrm{jets}}}
\newcommand{\ljets}{\ensuremath{\ell+\textrm{jets}}}
\newcommand{\dQ}{down-type quark}
\newcommand{\bQ}{$b$-quark}
\newcommand{\bjet}{$b$-jet}
\newcommand{\btag}{$b$-tag}

\def\et{\ensuremath{E_{\mathrm{T}}}} 
\def\ifb{\mbox{fb$^{-1}$}}
\newcommand{\etmiss}{\ensuremath{\et^{\textrm{miss}}}}
\newcommand{\transmass}{\ensuremath{m_{\textrm{T}}(W)}}

\begin{document}
\title{First Measurements of Spin Correlation Using Semi-leptonic $t\bar{t}$ Events at ATLAS}

\author{Boris Lemmer\\ On behalf of the ATLAS Collaboration}

\address{II. Physikalisches Institut, Georg-August-Universit\"at G\"ottingen}

\ead{boris.lemmer@cern.ch}

\begin{abstract}
The top quark decays before it hadronizes. Before its spin state can be changed in a process of strong interaction, it is directly transferred to the top quark decay products.
The top quark spin can be deduced by studying angular distributions of the decay products. The Standard Model predicts the top/anti-top quark (\ttbar) pairs to have correlated spins. The degree is sensitive to the top-quark spin itself and the production mechanisms of the top quark. Measuring the spin correlation allows to test the predictions. New physics effects can be reflected in deviations from the prediction. The measurement of the spin correlation of \ttbar\ pairs, produced at the LHC with a \com\ energy of $\sqrt{s} = 7\,\TeV$ and reconstructed with the ATLAS detector, is presented. The dataset corresponds to an integrated luminosity of \intlumi. 
\ttbar\ pairs are reconstructed in the \ljets\ channel using a kinematic likelihood fit offering the identification of light up- and down-type quarks from the $t \rightarrow bW \rightarrow bq\bar{q}'$ decay. The spin correlation is measured via the distribution of the azimuthal angle \dphi\ between two top quark spin analyzers in the laboratory frame. It is expressed as the degree of \ttbar\ spin correlation predicted by the Standard Model, \fsm. The result of $\fsm = 1.12 \pm 0.11\,\text{(stat.)} \pm 0.22\,\text{(syst.)}$
is consistent with the Standard Model prediction of $\fsm = 1.0$. 
\end{abstract}

\section{Spin Correlation in \ttbar\ Pairs}
The top quark as heaviest of all known quarks decays with a lifetime of $3.29^{+0.90}_{-0.63} \cdot 10^{-25}\,\textrm{s}$ \cite{lifetime}. This is at least one order of magnitude shorter than the timescale of hadronization and depolarization and allows to deduce the top quark's spin information via the measurement of angular distributions of its decay products \cite{hadtime1, hadtime2}. Within the Standard Model (SM), top quarks are produced almost unpolarized via the strong interaction. However, when being produced in pairs their spins are predicted to be correlated. The degree of observed correlation depends on the production and decay mechanisms of \ttbar\ pairs. Deviations from the SM prediction can give hints for physics beyond the Standard Model, like the production via a high-mass $Z'$ boson \cite{zprime, zprime2}, a heavy Higgs boson \cite{heavyhiggs} or the decay of a top quark into a bottom quark and a charged Higgs boson \cite{chargedhiggs1, chargedhiggs2}. 

\ttbar\ spin correlations have been measured by the CDF and D0 collaborations using angular distributions of the $t\bar{t}$ decay products \cite{CDFspin, D0spin}. The D0 collaboration reported a first evidence for non-vanishing spin correlation using a matrix-element approach in the \ljets\ channel \cite{D0MEM}. At the LHC \cite{LHC} the CMS collaboration measured the \ttbar\ spin correlation using the difference of azimuthal angles between the two charged leptons in the laboratory frame, \dphi, in dileptonic \ttbar\ events \cite{CMSspin}. This observable is sensitive to \ttbar\ spin correlation arising from gluon fusion production. The fact that this mechanism is dominant at the LHC and that \dphi\ can be measured without any event reconstruction (next to the identification of spin analyzers) makes it the preferred observable at the LHC. ATLAS \cite{ATLAS} reported the observation of \ttbar\ spin correlation using 2.1 \ifb\ of LHC data taken at a \com\ energy of 7 \TeV\ \cite{ATLASobservation} and exploited several other observables sensitive to \ttbar\ spin correlations using the full 7 \TeV\ dataset \cite{ATLASfull}. In the same publication, the first measurement of \ttbar\ spin correlation in the \ljets\ channel at the LHC is reported. This measurement is outlined in the following.

All so far measured values of the \ttbar\ spin correlation, which are complementary between Tevatron and LHC measurements due to the different production mechanisms and \com\ energies, agree with the SM predictions. 

\section{Dataset and Selection}
For the presented measurement of \ttbar\ spin correlation the full 2011 dataset, corresponding to an integrated luminosity of \intlumi and taken by the ATLAS detector, is used. The event selection requires one isolated lepton (electron or muon) and at least four jets of which at least one has to be identified as \bjet\ using a neural network algorithm with a tagging efficiency of 70\%. For the suppression of multijet background additional requirements on the missing transverse momentum \etmiss\ and the transverse $W$ boson mass \transmass\ are made ($\etmiss > 30 \GeV$ and $\transmass > 60 \GeV$ in the \ejets\ channel, $\etmiss > 20 \GeV$ and $\etmiss + \transmass > 60 \GeV$ in the \mujets\ channel). 

\section{Reconstruction of Hadronic Spin Analyzers}
Several top quark decay products can be utilized as spin analyzers to get information about the top quark spin via the measurement of their angular distributions. The sensitivity of an analyzer to the top quark spin is expressed by the spin analyzing power $\alpha$ with $-1 \leq \alpha \leq 1$. The charged leptons $\ell$ are preferred due to $\alpha_{\ell}^{\textrm{NLO}} = 0.998$. As the corresponding weak isospin particle from hadronically decaying $W$ bosons, the \dQ\ with  $\alpha_{down-type~quark}^{\textrm{NLO}} = 0.930$ is the most powerful hadronic top spin analyzer. However, as it is hard to reconstruct the \dQ, the \bQ\ from the top quark decay is a good alternative. The lower spin analyzing power of $\alpha_{b-\textrm{quark}}^{\textrm{NLO}} = -0.390$  is compensated by its higher reconstruction efficiency. The \ttbar\ events are reconstructed using a kinematic likelihood fitter (KLFitter) \cite{KLFitter}. For each possible quark-to-jet mapping it maximizes a likelihood function including constraints from the masses of the top quarks and the $W$ bosons as well as transfer functions describing the detector measurements of the energies of all jets and leptons from the \ttbar\ decay. To get additional discrimination power to separate the light up- and down-type quarks from the hadronically decaying $W$ bosons, two-dimensional probability distributions of the transverse momentum versus the weight of the \btag ging algorithm are multiplied to the maximized likelihood. The first quantity is different for light up- and down-type quarks due to the V-A structure of the weak decay vertex, while the second one can separate charm/strange quark pairs. For each event, the event probabilities of all possible $4!=24$ permutations are normalized and the permutation with the highest event probability is chosen. Figure \ref{fig:eventprobability} shows the event probability distribution of all reconstructed events. 
\begin{figure}[h]
\includegraphics[width=0.45\textwidth]{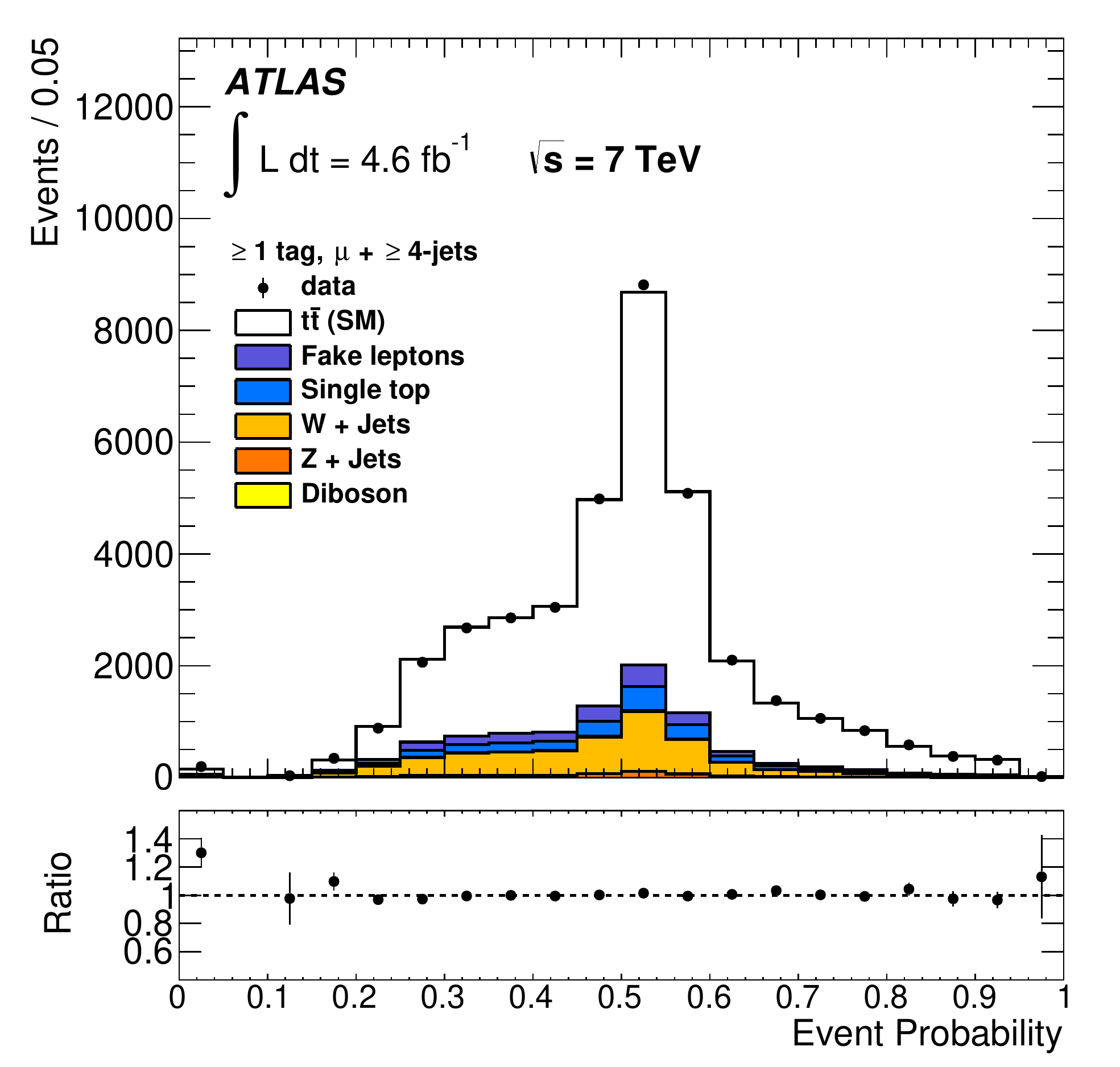}\hspace{2pc}%
\begin{minipage}[b]{0.45\textwidth}\caption{\label{fig:eventprobability} Distribution of the KLFitter event probabilities of the reconstructed \ttbar\ events \cite{ATLASfull}.}
\end{minipage}
\end{figure}
The peak at 0.5 illustrates the events where two permutations are undistinguishable, namely those where no charm/strange but up/down quarks are produced with a similar transverse momentum. For all events with event probabilities above 0.5 a good separation between light up- down-type quarks is reached. No cut on the event probability is applied.

\section{Fit Results and Uncertainties}
A binned likelihood fit is performed to fit signal and background contributions to the \dphi\ distributions  in data using the \dQ\ (\dphidQ) and the \bQ\ (\dphibQ). Priors corresponding to the normalization uncertainties were applied to the background and contribute to the statistical uncertainty. The \ttbar\ signal is mixed out of two samples, corresponding to uncorrelated \ttbar\ pairs ($n_{\ttbar}^{\textrm{unc.}}$) and \ttbar\ pairs with a spin correlation as in the SM ($n_{\ttbar}^{\textrm{SM}}$). The linearity of the spin correlation allows this mixing, which is determined by a parameter $s$ scaling the predicted \ttbar\ cross section to the measured one and the parameter \fsm:
\begin{align}
n_{\ttbar}^{\textrm{meas.}} = s \cdot  \left( \fsm \cdot n_{\ttbar}^{\textrm{SM}} + \left( 1- \fsm \right) \cdot n_{\ttbar}^{\textrm{unc.}} \right)
\end{align}
The fit is performed in channels of different jet multiplicities (4, $\geq 5$), numbers of \btag ged jets (1, $\geq 2$) and lepton flavours ($e$, $\mu$). If possible, systematic uncertainties are included in the fit via nuisance parameters. The other uncertainties are evaluated via ensemble tests. The fact that the spin analyzing powers of the \dQ\ and the \bQ\ have opposite signs cause the shape changes from SM spin correlation to uncorrelated \ttbar\ pairs to go into opposite directions. Combining the two analyzers hence reduces the effect of systematic uncertainties to a large extent. It is found that the shape is largely affected by kinematic mismodeling, in particular for the case of top quark $p_T$. The result of the combined fit, $f_{\rm SM} = 1.12 \pm 0.11~\textrm{(stat.)} \pm 0.22~\textrm{(syst)}$, is shown in Figure \ref{fig:result_dQ} and \ref{fig:result_bQ} and the uncertainties are listed in Table \ref{tab:uncertainties}. 

\begin{table}
\caption{\label{tab:uncertainties} Systematic uncertainties on the measured value of \fsm\ \cite{ATLASfull}.}
\begin{center}
\begin{tabular}{ll}
\br
  Detector modeling (nuisance parameters)                  & $\pm$ 0.09   \\
  Detector modeling (ensemble tests)                  & $\pm$ 0.02   \\

  Renormalization/factorization scale 		  &   $\pm$ 0.06  \\
  Parton shower and fragmentation     		  &   $\pm$ 0.16  \\
  ISR/FSR	                           	  &  $\pm$ 0.07  \\
  Underlying event                    		  &   $\pm$ 0.05  \\
  Color reconnection                 		  &   $\pm$ 0.01  \\
  PDF uncertainty                         	  &   $\pm$ 0.02  \\
  MC statistics                           	  &   $\pm$ 0.05  \\
  Top $p_{\rm T}$ reweighting                     &   $\pm$ 0.02 \\
\mr
  Total systematic uncertainty              	  &   $\pm$ 0.22	\\
  Data statistics                         	  &   $\pm$ 0.11     \\
\br
\end{tabular}
\end{center}
\end{table}

The measured results are in good agreement with the SM and with the results measured in the dilepton channel \cite{ATLASfull}.

\begin{figure}[h]
\begin{center}
\begin{minipage}{0.45\textwidth}
\includegraphics[width=\textwidth]{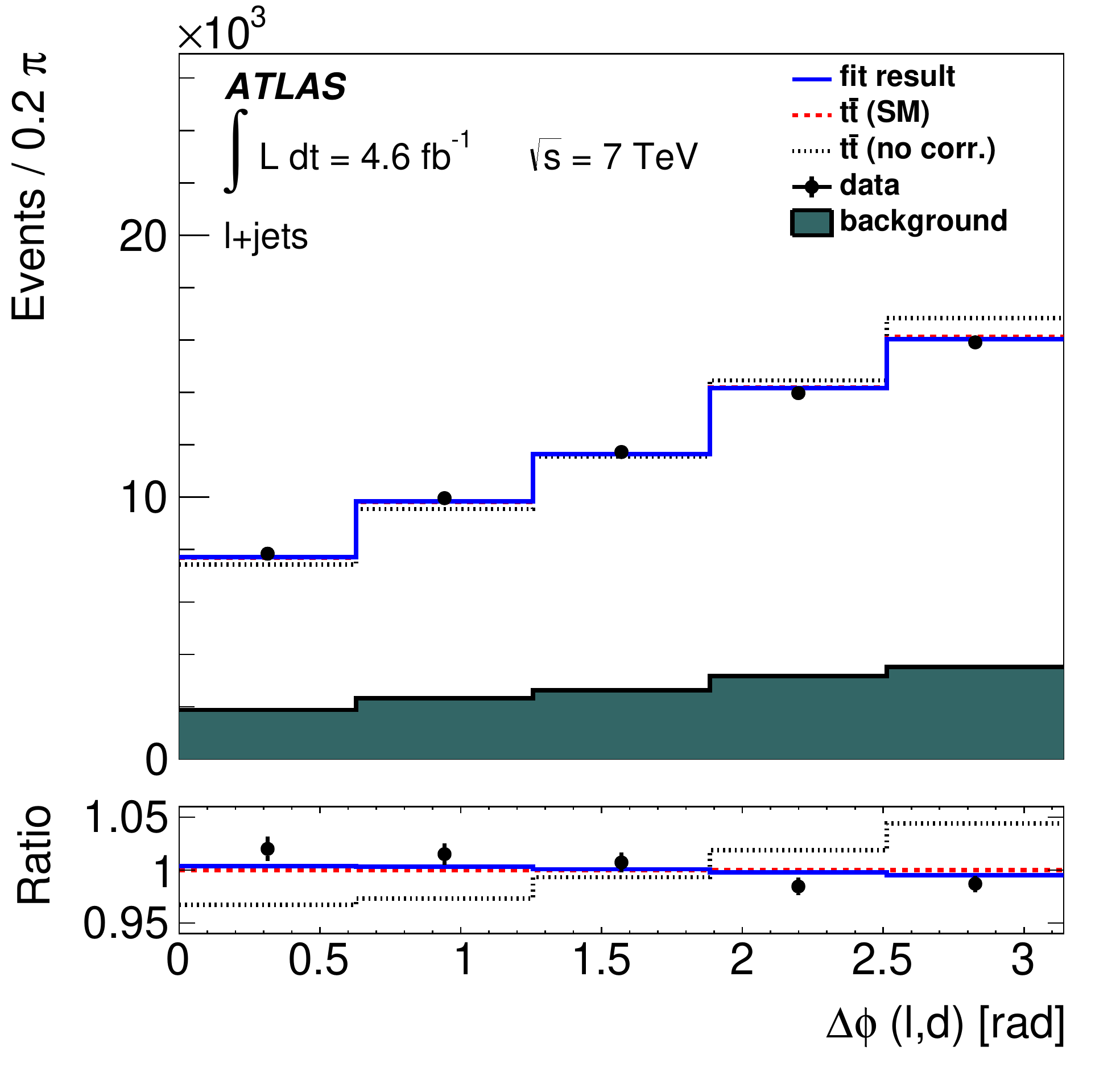}
\caption{\label{fig:result_dQ}Fitted distribution of \dphidQ\ together with the distribution of uncorrelated \ttbar\ pairs and \ttbar\ pairs with SM spin correlation \cite{ATLASfull}.}
\end{minipage}\hspace{2pc}%
\begin{minipage}{0.45\textwidth}
\includegraphics[width=\textwidth]{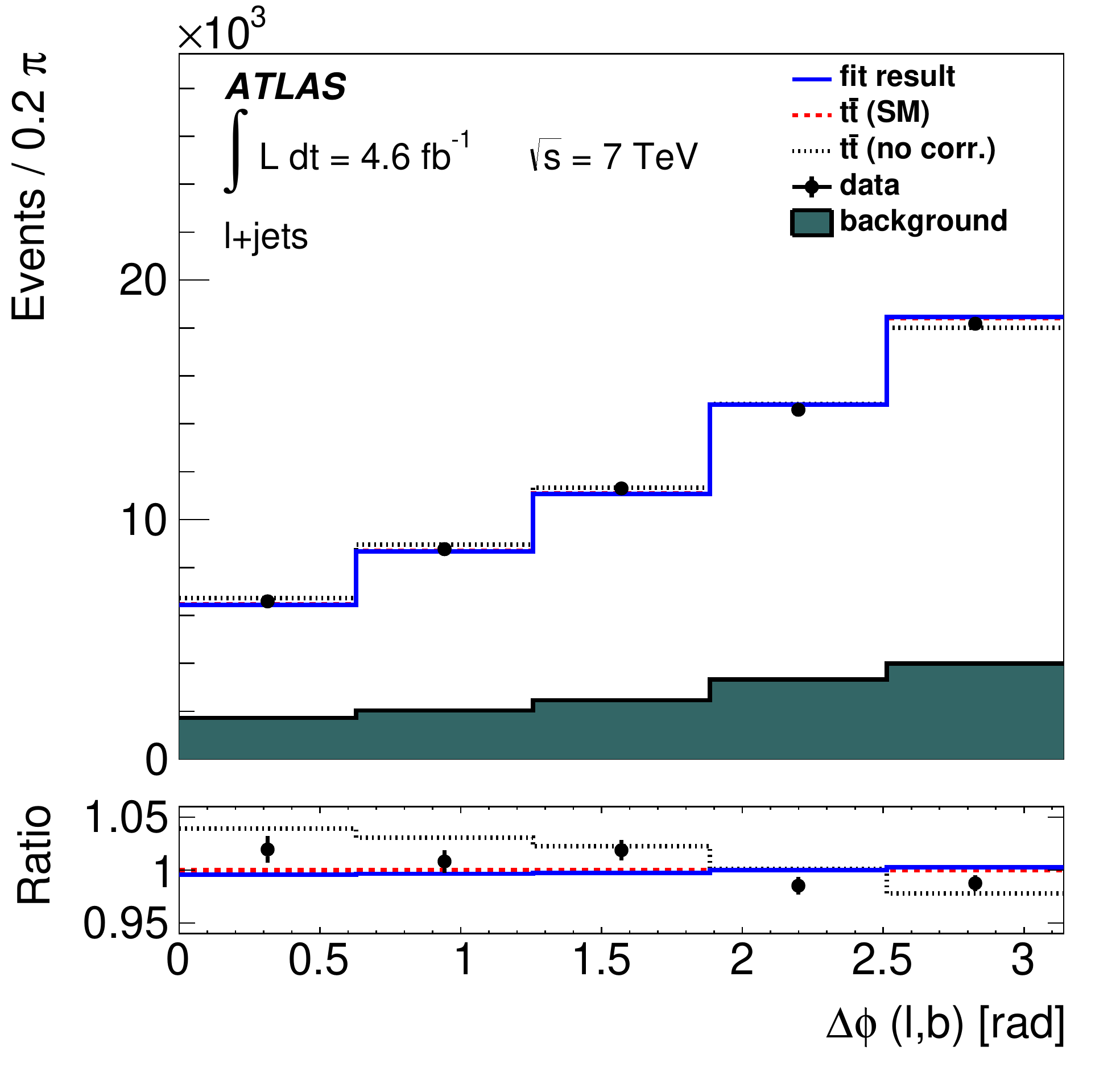}
\caption{\label{fig:result_bQ}Fitted distribution of \dphibQ\ together with the distribution of uncorrelated \ttbar\ pairs and \ttbar\ pairs with SM spin correlation \cite{ATLASfull}.}
\end{minipage} 
\end{center}
\end{figure}

\end{document}